\documentclass[conference]{IEEEtran}
\usepackage{amsmath}
\usepackage{eqparbox}
\usepackage{amssymb}
\usepackage{bm}
\usepackage{mathtools,mathabx}
\usepackage{amstext}

\usepackage{graphicx}
\usepackage{color}
\usepackage{booktabs}
\usepackage{longtable}
\usepackage{multicol}
\usepackage{amsfonts}
\usepackage{dsfont}
\usepackage{array}
\usepackage[most]{tcolorbox}
\usepackage[ruled]{algorithm}
\usepackage{algorithmic}
\usepackage{cases}
\usepackage{pgfplots}
\pgfplotsset{compat=newest}
\usetikzlibrary{spy}

\usepackage{amsthm}
\usepackage{caption}
\DeclareCaptionLabelSeparator{periodspace}{.~}
 \usepackage{float}
\usepackage{subcaption}

\usepackage{cite}
\usepackage{epstopdf}

\newcommand{\gf}[1]{\textcolor{black}{{#1}}}

\newcommand{\mx}[1]{\mathbf{#1}}
\newcommand{\bs}[1]{\boldsymbol{#1}}
\renewcommand{\triangleq}{\mathbin{\setstackgap{S}{0pt}\stackMath\Shortstack{\smalltriangleup\\ =}}}
\usepackage[usestackEOL]{stackengine} 

\def\pgfplotsinvokeiflessthan#1#2#3#4{%
    \pgfkeysvalueof{/pgfplots/iflessthan/.@cmd}{#1}{#2}{#3}{#4}\pgfeov
}%
\def\pgfplotsmulticmpthree#1#2#3#4#5#6\do#7#8{%
    \pgfplotsset{float <}%
    \pgfplotsinvokeiflessthan{#1}{#4}{%
        #7%
    }{%
        \pgfplotsinvokeiflessthan{#4}{#1}{%
            #8%
        }{%
            \pgfplotsset{float <}%
            \pgfplotsinvokeiflessthan{#2}{#5}{%
                #7%
            }{%
                \pgfplotsinvokeiflessthan{#5}{#2}{%
                    #8%
                }{%
                    \pgfplotsset{float <}%
                    \pgfplotsinvokeiflessthan{#3}{#6}{%
                        #7%
                    }{%
                        #8%
                    }%
                }%
            }%
        }%
    }%
}%

\theoremstyle{plain}

\theoremstyle{definition}

\theoremstyle{remark}

\newcommand{\RN}[1]{%
\textup{\uppercase\expandafter{\romannumeral#1}}%
}

\usepackage{standalone}
\graphicspath{{./Figures/}}

\definecolor{azure}{rgb}{0.0, 0.5, 1.0}
\definecolor{redjigar}{rgb}{0.9, 0.01, 0.1}
\definecolor{chestnut}{rgb}{0.8, 0.36, 0.36}
\definecolor{airforceblue}{rgb}{0.36, 0.54, 0.66}
\definecolor{cadmiumorange}{rgb}{0.93, 0.53, 0.18}
\definecolor{bleudefrance}{rgb}{0.19, 0.55, 0.91}
\definecolor{carolinablue}{rgb}{0.6, 0.73, 0.89}
\definecolor{blue(ncs)}{rgb}{0.0, 0.53, 0.74}
\definecolor{dodgerblue}{rgb}{0.12, 0.56, 1.0}
\definecolor{cssgreen}{rgb}{0.0, 0.5, 0.0}
\definecolor{cadmiumgreen}{rgb}{0.0, 0.42, 0.24}
\definecolor{cadmiumorange}{rgb}{0.93, 0.53, 0.18}
\definecolor{amaranth}{rgb}{0.9, 0.17, 0.31}
\definecolor{bluegray}{rgb}{0.4, 0.6, 0.8}
\definecolor{cadmiumgreen}{rgb}{0.0, 0.42, 0.24}

\begin{document}
%
\title{One Target, Many Views: Multi-User Fusion for Collaborative Uplink ISAC}

\author{\IEEEauthorblockN{Sajad Daei$^\dagger$, Gabor Fodor$^{\dagger\star}$, Mikael Skoglund$^\dagger$}
 \IEEEauthorblockA{$^\dagger$School of Electrical Engineering and Computer Science, KTH Royal Institute of Technology, Stockholm, Sweden\\
$^\star$Ericsson Research, Sweden\\ 
Email: \{sajado, gaborf, skoglund\}@kth.se}
}

\maketitle
    
\begin{abstract}

We propose a novel pilot-free multi-user uplink framework for integrated sensing and communication (ISAC) in mm-wave networks, where single-antenna users transmit \gf{orthogonal frequency division multiplexing} signals without dedicated pilots. The \gf{base station} exploits the spatial and velocity diversities of users to simultaneously decode messages and detect targets, transforming user transmissions into a powerful sensing tool. 
Each user's signal, structured by a known codebook, propagates through a sparse multi-path channel with shared moving targets and user-specific scatterers. 
Notably, common targets induce distinct delay–Doppler–angle signatures, while stationary scatterers cluster in parameter space.
We formulate {the} joint multi-path parameter estimation and data decoding as a 3D super-resolution problem, extracting delays, Doppler shifts, and angles-of-arrival via atomic norm minimization, efficiently solved using semidefinite programming. A core innovation is multi-user fusion, where diverse user observations are collaboratively combined to enhance sensing and decoding. This approach improves robustness and integrates multi-user perspectives into a unified estimation framework, enabling high-resolution sensing and reliable communication.
Numerical results show that the proposed framework significantly enhances both target estimation and communication performance, highlighting its potential for next-generation ISAC systems.
\end{abstract}

%
%
\makeatletter{\renewcommand*{\@makefnmark}{}
\footnotetext{This work is supported in part by Digital Futures. G.Fodor is also supported by the Swedish Strategic Research (SSF) grant for the FUS21-0004 SAICOM project and the 6G-Multiband Wireless and Optical Signaling for Integrated Communications, Sensing and Localization (6G-MUSICAL) EU Horizon 2023 project, funded by the EU, Project ID: 101139176.
S.Daei is supported by Digital Futures Project STACEY. }
\makeatother}

%
\IEEEpeerreviewmaketitle

\section{Introduction}\label{section1}

The rapid evolution of wireless networks demands systems that not only communicate but also sense their environment. \gf{Integrated sensing and communication (ISAC)} has emerged as a key enabler of this vision, allowing the same resources to be used for both data transmission and target detection \cite{liu2020joint,hassanien2016signaling,survey1,liu2022integrated,liu2022survey,collab_sense1,colab_sense2,iman1,liu2023seventy,liu2025sensing,meng2024cooperative}. This dual functionality leads to substantial gains in spectral efficiency, hardware utilization, and energy consumption.

While ISAC research has extensively focused on single-user or downlink-based architectures, multi-user uplink scenarios at millimeter-wave (mm-wave) frequencies present unique opportunities and challenges \cite{rappaport2015millimeter,liu2020joint,liu2023seventy,liu2025sensing}. 
In such settings, each user's transmission acts as both a communication signal and a radar illumination source, enhancing sensing capabilities but also introducing potential interference in data decoding. 
Effectively exploiting this multi-user diversity requires novel signal processing techniques that seamlessly 
integrate communication and sensing.

In this paper, we propose a pilot-free multi-user ISAC uplink framework, in which multiple single-antenna users transmit orthogonal frequency division multiplexing (OFDM) signals to a base station (BS) equipped with ISAC functionality. Unlike traditional multi-user uplink architectures that rely on explicit pilots, our approach leverages the inherent sparsity of the mm-wave propagation environment to jointly estimate delay, Doppler, and \gf{angle-of-arrival (AoA)} parameters, while simultaneously decoding user data. The system operates in a blind setting, where the BS has only knowledge of users' structured codebooks and must extract all relevant parameters without direct pilot assistance.

A key distinguishing feature of our framework is its exploitation of multi-user diversity in both the spatial and velocity domains. Conventional ISAC uplink approaches often overlook these forms of diversity\cite{collab_sense1,meng2024cooperative,colab_sense2}. 
In contrast, we show that the spatial separation of users provides multiple perspectives of common targets, while velocity differences introduce distinct Doppler shifts, enriching both sensing and decoding capabilities. 
This dual-diversity effect is particularly beneficial for distinguishing moving targets from stationary 
scatterers and mitigating user interference.

To solve the joint sensing and communication problem, we formulate a \gf{3-dimensional (3D)} super-resolution estimation framework that extracts delay, Doppler, and \gf{AoA} parameters. 
We achieve this by employing atomic norm minimization, efficiently solved via semidefinite programming (SDP) \cite{candes2014towards,tang2013compressed,fernandez2016super,valiulahi2019two,sayyari2020blind,bayat2020separating,daei_icassp,daei2024timely,daei2023off-the-grid}. Our method does not rely on pilot signals; instead, it directly recovers parameters from structured transmissions by leveraging the sparse representation of the mm-wave channel in the delay-Doppler-angle domain.

Unlike conventional multi-user uplink systems, where adding users can degrade decoding performance, our ISAC approach transforms additional users into an asset. Each user provides a unique and distinct "view" of the environment, 
enhancing common target localization, velocity estimation, and communication channel knowledge. 
This is accomplished through a collaborative multi-user fusion process, where users’ independent perspectives of the same target are aggregated to improve parameter estimation accuracy and message decoding reliability. 
Depending on whether users share or possess distinct codebooks, we introduce multiple fusion strategies, including weighted averaging, pointwise averaging, and alignment of dual polynomials. 
These fusion techniques mitigate noise, enhance resolution, and yield significant improvements in both 
sensing accuracy and data decoding performance.

\subsection{Contributions}
The key contributions of this paper are as follows:
\begin{enumerate}
   \item \textbf{Blind Multi‐User ISAC Uplink:} We develop a pilot-free ISAC uplink framework that reformulates multi-user OFDM data transmission as a 3D super-resolution problem. Our approach jointly exploits user transmissions for both target sensing and data decoding, enabling the estimation of shared and user-specific scatterers.

   \item \textbf{Atomic Norm and SDP-Based Parameter Estimation:} We formulate two novel optimization frameworks to estimate the continuous-valued delay, Doppler, and \gf{AoA} parameters without relying on explicit pilot signals. The first approach leverages Vandermonde decomposition for direct continuous parameter recovery, while the second employs 3D dual polynomials for robust estimation. We demonstrate the effectiveness of both methods in achieving high-precision parameter recovery and robust message decoding.

    \item \textbf{Collaborative Dual Polynomial Techniques:} We propose collaborative estimation strategies that exploit multi-user diversity, including weighted averaging, pointwise averaging, and aggregated dual polynomial using codebook alignment. By processing multiple dual polynomials jointly, we enhance the target detection performance and improve the robustness in parameter estimation.

    \item \textbf{Exploiting Dual Diversity for Enhanced Estimation:} Unlike conventional uplink designs that treat user transmissions as interference, our framework leverages both spatial and Doppler diversity--arising from different user locations and velocities--to enhance parameter estimation. This dual-diversity effect enables sharper target localization and improved data decoding, even under noisy conditions.
\end{enumerate}

\textbf{Organization of the Paper:} 
The remainder of the paper is structured as follows. Section \ref{sec.model} introduces the uplink system model, the OFDM data transmissions, and the sparse channel representation. Section \ref{sec.proposed} outlines our atomic norm approach, explains the SDP problem, and describes how dual polynomials reveal the user and target parameters. Section \ref{sec.simulations} presents simulation results demonstrating both sensing and communication gains. Finally, Section \ref{sec.conclusion} provides the conclusions of the paper and future directions.

\textbf{Notations:}
Bold lower-case letters (e.g., \(\bm{x}\)) denote vectors, while bold upper-case letters (e.g., \(\bm{X}\)) denote matrices. The transpose and Hermitian (conjugate-transpose) of a matrix \(\bm{X}\) are denoted by \(\bm{X}^{\mathsf{T}}\) and \(\bm{X}^{\mathsf{H}}\), respectively. For a vector \(\bm{x}\in\mathbb{C}^n\) and a matrix \(\bm{X}\in\mathbb{C}^{n_1\times n_2}\), the \(\ell_2\) norm and Frobenius norm are denoted by \(\|\bm{x}\|_2\) and \(\|\bm{X}\|_{\mathrm{F}}\), respectively. We write \(\bm{X} \succeq \bm{0}\) to indicate that \(\bm{X}\) is positive semidefinite.
For any two matrices \(\bm{A}\) and \(\bm{B}\) (of compatible dimensions), the inner product is defined as
$\langle \bm{A}, \bm{B} \rangle \triangleq \mathrm{tr}\!\left(\bm{B}^{\mathsf{H}}\bm{A}\right).$
The canonical basis vector in \(\mathbb{R}^{N}\) is denoted by \(\bm{e}_n\), whose \(n\)th element is one and all other elements are zero. The imaginary part of a scalar \(x\) is denoted by \({\rm Im}(x)\), and \(x_n\) or \([x]_n\) represents the \(n\)th element of a vector \(\bm{x}\in\mathbb{C}^{N}\). The set of $\{1,..., R\}$ is denoted by $[R]$. The complex conjugate of a variable $x$ is shown by $x^\ast$.

\section{System Model and Problem Formulation}\label{sec.model}

We consider a multi-user uplink system at mm-wave frequencies, where \(r\) single-antenna ISAC-enabled users transmit OFDM waveforms to a BS equipped with a \gf{uniform linear array} of \(N_{r}\) antennas. 
Crucially, the system is \emph{pilot-free}: rather than using separate pilot signals, the BS exploits the data-bearing uplink transmissions for both message decoding and target sensing, thereby achieving greater spectral efficiency.

\subsection{OFDM Signaling and Key Parameters}

Each user \(i\) generates a continuous-time OFDM signal\cite{zheng2017super}:
$$
  x_i(t)
  \;=\;
  \sum_{p=0}^{P-1} x_{i,p}(t)
  \;=\;
  \sum_{p=0}^{P-1}\sum_{q=0}^{Q-1}
  g_{i,p}(q)\,{\rm e}^{\,j\,2\pi\,q\,\Delta f\,t}\,\phi\!\bigl(t - p\,\overline{T}\bigr).
$$

Here, \(p \in \{0,\ldots,P-1\}\) indexes each OFDM block (or packet), and \(q \in \{0,\ldots,Q-1\}\) indexes the subcarriers. The term \(g_{i,p}(q)\) denotes the data symbol of user \(i\) on subcarrier \(q\) in block \(p\). The function \(\phi(t)\) is nonzero only on \([-T_{\mathrm{cp}},\,T]\), effectively modeling one OFDM symbol of duration \(T\) augmented by a \gf{cyclic prefix} of length \(T_{\mathrm{cp}}\). The total block duration is \(\overline{T} = T + T_{\mathrm{cp}}\). The subcarrier spacing is \(\Delta f = 1/T\).

\subsection{Channel Model with Delay, Doppler, and AoA}
We model the channel between user \(i\) and the BS as a sparse sum of \(s_i\) propagation paths. These paths encompass stationary communication scatterers and possible radar targets. Some scatterers or targets may be \emph{shared} by multiple users (i.e., common), while others remain user-specific. Stationary scatterers, having negligible Doppler, often form clustered paths, whereas high-velocity targets appear as point-like reflectors with distinctly observable Doppler shifts.

For path \(\ell\) associated with user \(i\), we define a complex gain \(c_{\ell,i}\), a physical delay \(\overline{\tau}_{\ell,i}\), a Doppler shift \(\overline{\nu}_{\ell,i}\), and an \gf{angle of arrival (AoA)} \(\overline{\theta}_{\ell,i}\). Hence, the continuous-time channel impulse response for user \(i\) is
$$
  \mathbf{h}_i(t,\tau)
  \;=\;
  \sum_{\ell=1}^{s_i}
  c_{\ell,i}\,{\rm e}^{\,j\,2\pi\,\overline{\nu}_{\ell,i}\,t}
  \,\delta\!\bigl(\tau - \overline{\tau}_{\ell,i}\bigr)
  \,\mathbf{a}\!\bigl(\overline{\theta}_{\ell,i}\bigr)~\in  \mathbb{C}^{N_r \times 1},
$$
where \begin{equation}
\label{eq:ula_row_vector}
\mathbf{a}(\overline{\theta}_{\ell,i})
\;\triangleq\;
\bigl[\,1,
\ldots,
\;
{\rm e}^{\,j\,2\pi\,(N_{R}-1)\,\tfrac{d}{\lambda}\,\sin(\overline{\theta}_{\ell,i})}
\bigr]^\mathsf{T},
\end{equation} denotes the BS array steering vector corresponding to AoA \(\overline{\theta}_{\ell,i}\). 
Here, \(d\) represents the antenna spacing, \(\lambda\) is the wavelength, and $s_i$ denotes the number of communication scatterers or targets associated with the $i$-th user's channel.

\subsection{Received Signal Model}

Let \(\overline{\mathbf{y}}(t)\in\mathbb{C}^{N_{r}\times 1}\) represent the signal received at the BS. 
Convolving each user’s transmitted waveform \(x_i(\cdot)\) with its channel \(\mathbf{h}_i(\cdot)\) yields
$$
\overline{\mathbf{y}}(t)
\;=\;
\sum_{i=1}^R
\int 
\mathbf{h}_i(t,\tau)\,x_i\!\bigl(t-\tau\bigr)\,d\tau+\bs{\epsilon}(t),
$$
where $\bs{\epsilon}(t)$ is the additive noise vector in time and $R$ is the total number of users.
Expanding the summation over OFDM symbols and paths and focusing on the $r$-th antenna element gives\cite{zheng2017super}:
\begin{align}
\overline{y}_r(t)
&= \sum_{i=1}^R \sum_{p=0}^{P-1} \sum_{\ell=1}^{s_i}
c_{\ell,i}\,
x_{i,p}\bigl(t - \overline{\tau}_{\ell,i}\bigr)
\, e^{j\,2\pi\,\overline{\nu}_{\ell,i}\,t} \nonumber\\
&\hspace{3cm}
\times\, e^{j\,2\pi\,\tfrac{d}{\lambda}\,\sin(\overline{\theta}_{\ell,i})\,r}
+ \epsilon_r(t). \label{eq:received_signal}
\end{align}

Since \(\overline{\nu}_{\ell,i}\ll 1/\overline{T}\), we approximate the factor \({\rm e}^{\,j\,2\pi\,\overline{\nu}_{\ell,i}\,t}\) by a constant within each OFDM block. 
Consequently, \gf{the received signal at antenna $r$ can be written as:}
\begin{align}
    &\overline{y}_r(t)
\; \approx \;
\sum_{i=1}^R
\sum_{p=0}^{P-1}
\sum_{\ell=1}^{s_i}
c_{\ell,i}\,
x_{i,p}\!\bigl(t - \overline{\tau}_{\ell,i}\bigr)
\,{\rm e}^{\,j\,2\pi\,\overline{\nu}_{\ell,i}\,p\,\overline{T}}
\,\nonumber\\
&~~~~~~~~~~~~~~{\rm e}^{\,j\,2\pi\,\tfrac{d}{\lambda}\,\sin(\overline{\theta}_{\ell,i})\,r}+\epsilon_r(t).
\end{align}
\paragraph{Frequency-Domain Sampling}
To obtain subcarrier-domain measurements, we take the Fourier transform of 
\(\mathbf{y}(t)\) on each OFDM block (duration \(T\)) and sample at frequencies 
\(f_q = q\,\Delta f\), for \(q = 0, \dots, Q - 1\). 
Letting \(r \in \{0, \dots, N_r - 1\}\) index each antenna element, we define
\begin{equation}
\widetilde{y}_{p,r}\!\bigl(q\,\Delta f\bigr)
\;\triangleq\;
\int_{p\,\overline{T}}^{\,p\,\overline{T}+\,T}
\overline{y}_r(t)\,
e^{-\,j\,2\pi\,q\,\Delta f\,t}
\,dt.
\label{eq:freqDomainSample}
\end{equation}
Inserting the time‐domain expansion of \(\mathbf{y}(t)\) and letting 
\(g_{i,p}(q)\) denote the frequency-domain sample of user \(i\) on subcarrier 
\(q\) in the \(p\)-th block, we obtain:
\begin{align}
&\widetilde{y}_{p,r}\!\bigl(q\,\Delta f\bigr)
\;=\;
\sum_{i=1}^R
\sum_{\ell=1}^{s_i}
c_{\ell,i}\,
e^{-\,j\,2\pi\,\overline{\tau}_{\ell,i}\,\tfrac{q}{T}}
\,g_{i,p}(q)\,
e^{\,j\,2\pi\,\overline{\nu}_{\ell,i}\,p\,\overline{T}}
\,\nonumber\\
&~~~~~~~~~~~~~~~~~~e^{\,j\,2\pi\,\tfrac{d}{\lambda}\,\sin(\overline{\theta}_{\ell,i})\,r}+\widetilde{\epsilon}_{p,r}(q\Delta f),
\label{eq:freqDomainModel}
\end{align}
where $\widetilde{\epsilon}_{p,r}(q\Delta f)\triangleq \int_{p\,\overline{T}}^{\,p\,\overline{T}+\,T}
\bs{\epsilon}_r(t)\,
e^{-\,j\,2\pi\,q\,\Delta f\,t}
\,dt$ is the noise term in the frequency bin $q$, time bin $p$ and antenna element $r$ and is assumed to be zero mean with variance $\sigma^2$.
\paragraph{Unified Indexing of Time--Frequency--Space}
We now define a single index \(n\) to unify time, frequency, and antenna:
\begin{align}
    n &= p + Pq + PQr, \label{eq:unifiedIndex} \\
    &\quad p = 0,\ldots,P{-}1,\quad q = 0,\ldots,Q{-}1,\quad r = 0,\ldots,N_r{-}1. \nonumber
\end{align}

so that the total number of samples is 
\(L \triangleq P\,Q\,N_r\). 
In addition, we introduce the normalized parameters
\begin{equation}
\nu_{\ell,i} 
\;\triangleq\; 
\overline{\nu}_{\ell,i}\,\overline{T},
\quad
\tau_{\ell,i}
\;\triangleq\; 
\tfrac{\overline{\tau}_{\ell,i}}{T},
\quad
\theta_{\ell,i}
\;\triangleq\;
\tfrac{d}{\lambda}\,\sin\bigl(\overline{\theta}_{\ell,i}\bigr).
\label{eq:normalizedParams}
\end{equation}

The observed signal at the unified index \(n\) then becomes
\begin{align}
\label{eq:measurement2}
&y_n \triangleq\widetilde{y}_{p,r}\!\bigl(q\,\Delta f\bigr)
=\;
\sum_{i=1}^{R}
\sum_{\ell=1}^{s_i}
c_{\ell,i}\;
\mathbf{e}_n^\mathsf{T}\,
\mathbf{a}_{3D}\!\bigl(\zeta_{\ell,i}\bigr)
\,x_n^{(i)}+{\epsilon}_n,
\nonumber\\
&n=0,\dots,L-1,
\end{align}
where \(x_n^{(i)}\triangleq g_{i,p}(q)\) is the data symbol of user~\(i\) on subcarrier~\(q\) in block~\(p\). $\zeta\triangleq (\tau,\nu,\theta)$ shows the triple continuous variables. The vector \(\mathbf{a}(\zeta)\) has its \(n\)-th entry given by
$$
[\mathbf{a}_{3D}(\zeta)]_{n}
=
{\rm e}^{\,j\,2\pi\,\bigl(q\,\tau + p\,\nu + r\,\theta\bigr)},
$$
while \(\mathbf{e}_n\in\mathbb{R}^{L}\) denotes the \(n\)-th canonical basis vector. This observation model at the BS side captures the delay--Doppler--angle structure for each path within a single index scheme over subcarrier, OFDM block, and antenna array dimensions, providing a foundation for joint user data recovery and target parameter estimation in a pilot-free, multi-user ISAC uplink.

\section{Proposed method}\label{sec.proposed}
In this section, we first connect the signal model provided in \eqref{eq:measurement2} to a 3D super-resolution problem in Section \ref{sec:super_model}. Then, we propose two distinct methods for parameter recovery: the primal approach (detailed in Section~\ref{sec:primal_approach}) and the dual approach (described in Section~\ref{sec:dual_approach}). The primary advantage of the primal method is its relatively lower computational burden, as it directly recovers continuous parameters via a Vandermonde decomposition provided that $s_i\ll L-\max\{P,Q,N_r\}$. By contrast, the dual approach involves solving the dual problem to construct a dual polynomial, with the target parameters identified at the peaks of this polynomial. Although the dual method tends to be more robust in noisy scenarios, it comes at the cost of increased computational complexity.

\subsection{\gf{Signal Representation By Means of a Suitable Atomic Norm}}\label{sec:super_model}
The total number of unknown in \eqref{eq:measurement2} is $\mathcal{O}(LR+\sum_i s_i)$ and number of equations is only $L$ samples. The problem is inherently intractable. To make the problem trackable, we assume that the transmitted signal corresponding to user $i$ lies in a codebook $\mx{D}_i=[{\mx{d}^i}^{T}_0,...,{\mx{d}^i}^{\mathsf{T}}_{L-1}]^{\mathsf{T}}\in L\times k_i$ i.e., $x_n^i\triangleq {\mx{d}_n^{i}}^{\mathsf{T}} \mx{f}_i$. Here $f_i\in\mathbb{C}^{k_i\times 1}$ is the message transmitted by user $i$. 

The received signal model \eqref{eq:measurement2} can then be rewritten as:
\begin{align}\label{eq:mesaurement3}
   {y}_n= \sum_{i=1}^R \langle \mx{X}_i, \mx{B}^i_n \rangle+{\epsilon}_n,~~n=0,..., L-1,
\end{align}
where $\mx{X}_i\triangleq\sum_{\ell=1}^{s_i} c_{\ell,i} \mx{f}_i \mx{a}_{\rm 3D}^{\mathsf{T}}(\zeta_{\ell})$ and $\mx{B}^i_n\triangleq \mx{d}_n^i \mx{e}_n^{\mathsf{T}}$.

To jointly capture the sparsity feature of time-freq-space shifts, we define the following atomic norm minimization:
We thereby define the following atomic norm \cite{chandrasekaran2012convex}:
\begin{align}
\label{eq.atomic_def}
\|\mx{X}_i\|_{\mathcal{A}^{3D}_i}
&\triangleq \inf_{\{c_{\ell}, \zeta_{\ell}\}} 
\Bigg\{ \sum_{\ell} |c_{\ell}| \|\mx{f}_i\|_2~: \nonumber\\
&\qquad\quad \mx{X}_i = \sum_{\ell} c_{\ell}\, \mx{f}_i\, \mx{a}_{\rm 3D}(\zeta_{\ell})^{\mathsf{T}} 
\in \mathbb{C}^{k_i \times L} \Bigg\}
\end{align}
 associated with the atoms
 \begin{align}\label{eq.atoms}
 \mathcal{A}_i^{3D}=\big\{\mx{f}_i\mx{a}_{3D}(\zeta)^{\mathsf{T}}: \zeta\in{[0,1)}^3, \|\mx{f}_i\|_2=1, \mx{f}_i    \in\mathbb{C}^{k_i\times 1} \big\},
 \end{align}
 where $i\in [R]$. The atomic norm $\|\mx{X}_i\|_{\mathcal{A}_i}$ can be regarded as the best convex alternative for the smallest number of atoms $\mathcal{A}^{3D}_i$ needed to represent a signal $\mx{X}_i$. Hence, we are interested in recovering the matrices $\mx{X}_1,...,\mx{X}_R$ by encouraging their atomic sparsity via solving
\begin{align}\label{eq.primalprob_noisy}
 &\min_{\widetilde{\mx{y}}, (\mx{Z}_i)_{i=1}^R} ~\sum_{i=1}^{R}\|\mx{Z}_i\|_{\mathcal{A}^{3D}_i},\\
& {\rm s.t.}~\widetilde{{y}}_n={y}_{n}-\sum_{i=1}\langle \mx{Z}_i, \mx{B}_n^i \rangle, n=0,..., L-1 ,~
 \|\widetilde{\mx{y}}\|_2\le \eta.	
 \end{align} 
where $\eta$ is an upper-bound of $\|\bs{\epsilon}\|_2$.

\subsection{Primal Approach Via Vandermonde Decomposition}\label{sec:primal_approach}

\subsubsection{3‐Level Toeplitz Matrix}
Before stating our main result, we need to define the 3D Toeplitz matrix that is employed in our approach.
Let \(\mathbf{V}_i\in\mathbb{C}^{(2P-1) \times (2Q-1) \times (2N_r-1)}\) be a tensor whose entries are denoted by: 
\begin{align}
    V_i(k_1,k_2,k_3),\quad \\
k_1 \in \{-(P-1),\ldots,P-1\}, \\
k_2 \in \{-(Q-1),\ldots,Q-1\},\\
k_3 \in \{-(N_r-1),\ldots,N_r-1\}.
\end{align}

The corresponding three-level (3L) Toeplitz matrix \(T_{\rm 3L}(\mathbf{V}_i)\in\mathbb{C}^{L \times L}\) is constructed so that its entries depend only on the differences between the indices. Specifically, by indexing the rows and columns of \(T_{\rm 3L}(\mathbf{V}_i)\) by multi-indices 
\[
\mathbf{m}=(m_1, m_2, m_3) \quad \text{and} \quad \mathbf{n}=(n_1, n_2, n_3),
\]
with \(m_1,n_1 \in \{0,\ldots,P-1\}\), \(m_2,n_2 \in \{0,\ldots,Q-1\}\), \(m_3,n_3 \in \{0,\ldots,N_r-1\}\), the \((\mathbf{m},\mathbf{n})\)-th entry is defined as
\begin{equation}
\label{eq:3D_toeplitz}
\left[ T_{\rm 3L}(\mathbf{V}_i) \right]_{\mathbf{m},\mathbf{n}}
\;=\;
V_i\Bigl(m_1-n_1,\; m_2-n_2,\; m_3-n_3\Bigr).
\end{equation}

This construction yields a 3-level Toeplitz matrix because the entry depends solely on the differences \(m_j - n_j\) for \(j=1,2,3\). Such a structure naturally extends the conventional Toeplitz property for one dimension to three dimensions, ensuring that the matrix has a consistent block structure along each of the three indices.

\subsubsection{SDP Relaxation of the Primal Problem \eqref{eq.primalprob_noisy} }

Building on multidimensional line spectral estimation results 
\cite{chandrasekaran2012convex,dumitrescu2017positive,yang2016vandermonde}, 
we can express $\|\mathbf{Z}_i\|_{\mathcal{A}^{3D}_i}$ in \eqref{eq.primalprob_noisy} as an SDP.  
Introduce variables $t_i>0$ and $\mathbf{V}_i\in\mathbb{C}^{(2P-1)\times (2Q-1)\times (2N_r-1)}$,
\gf{and} let $T_{\rm 3L}(\mathbf{V}_i)\in\mathbb{C}^{L\times L}$ 
be the corresponding \emph{3L Toeplitz} matrix. The 3D atomic norm can be alternatively stated as the solution to the following problem which is an adapted 3D version of \cite[Proposition 1]{bayat2020separating}, given by:
\begin{align}
\label{eq:3DatomicSDP}
&\|\mathbf{Z}_i\|_{\mathcal{A}^{3D}_i}
~=\;
\min_{\substack{t_i>0 \\ \mathbf{V}_i}}
~
\tfrac{1}{2\,L}
\,\mathrm{Tr}\!\bigl(T_{\rm 3L}(\mathbf{V}_i)\bigr)
~+~
\tfrac{1}{2}{\rm tr}(\mx{W}_i)
\nonumber\\
&\text{subject to}\begin{bmatrix}
T_{\rm 3L}(\mathbf{V}_i) & \mathbf{Z}_i\\
\mathbf{Z}_i^{\mathsf{H}} & \mx{W}_i
\end{bmatrix}
\succeq
\mathbf{0},
\end{align}
 
Substituting \eqref{eq:3DatomicSDP} into \eqref{eq.primalprob_noisy} 
yields the final \emph{3D super‐resolution} problem:
\begin{align}
\label{eq:3DfinalSDP}
&\min_{\widetilde{\mathbf{y}},\,\{\mathbf{Z}_i,\mathbf{V}_i,t_i\}}
\quad
\sum_{i=1}^R
\Bigl[
\tfrac{1}{2\,L}\,\mathrm{Tr}\bigl(T_{\rm 3L}(\mathbf{V}_i)\bigr)
~+~
\tfrac{1}{2}{\rm tr}(\mx{W}_i)
\Bigr],
\\
&\text{subject to}
\quad
\widetilde{y}_n
~=~
y_n
~-\!\sum_{i=1}^R
\langle
\mathbf{Z}_i,\,\mathbf{B}_n^i
\rangle,
\quad
\|\widetilde{\mathbf{y}}\|_2
\le
\eta,
\nonumber
\\
&\qquad\qquad
\begin{bmatrix}
T_{\rm 3L}(\mathbf{V}_i) & \mathbf{Z}_i\\
\mathbf{Z}_i^{\mathsf{H}} & \mx{W}_i
\end{bmatrix}
\succeq
\mathbf{0},
\;\;
i=1,\dots,R.
\nonumber
\end{align}
This SDP enforces both the measurement constraints 
and the 3L Toeplitz property, capturing the 
delay-Doppler-angle sparsity in $\mathbf{X}_i$. The problem \eqref{eq:3DfinalSDP} can be solved by off-the-shell SDP solvers in CVX \cite{grant2014cvx}. This provides an estimate of the 3L Toeplitz matrix $T_{\rm 3L}(\mx{V}_i),i=1..., R$ which has the information of 3D continuous parameters of all users. In the next section, we provide a decomposition method to estimate the 3D parameters from this Toeplitz matrix.

 \subsubsection{Vandermonde Decomposition of a 3L Toeplitz Matrix}

The 3L Toeplitz matrix $T_{\rm 3L}(\mx{V}_i)$ can be decomposed as a weighted sum of rank-one matrices as follows:
\begin{equation}
     {T}_{3L}(\mx{V}_i) = \sum_{\ell=1}^{s_i} p_{\ell} \, \mathbf{a}_{3D}(\zeta_{\ell}) \mathbf{a}_{3D}(\zeta_{\ell})^{\mathsf{H}},
    \label{eq:decomp}
\end{equation}
where $p_{\ell}$ is the power (or amplitude) associated with the $\ell$-th 3D parameter $\zeta_{\ell}$. In what follows, we propose a method based on matrix pencil and 3D pairing \cite{roy1989esprit} called ${\rm MaPP}_{\rm 3D}$.

\subsubsection{ ${\rm MaPP}_{\rm 3D}$ for Vandermonde Decomposition}\label{sec:mapp_3d}
${\rm MaPP}_{\rm 3D}$ is composed of four steps which is explained below:  
    \textbf{Eigen-Decomposition:} The algorithm begins by computing the eigen-decomposition of the symmetrized matrix $\tfrac{ {T}_{3L}(\mx{V}_i)+ {T}_{3L}(\mx{V}_i)^{\mathsf{H}}}{2}$. By selecting the $s_i$ dominant eigenvalues and their corresponding eigenvectors, a matrix $\mathbf{U}_i\in\mathbb{C}^{L\times s_i}$ is formed:
    \begin{equation}
  {T}_{3L}(\mx{V}_i) = \mathbf{U}_i \bs{\Lambda}_i\mx{U}_i^{\mathsf{H}}.
    \end{equation}
    This step effectively isolates the signal subspace corresponding to the delay-Doppler-angle components.

    \textbf{Parameter Estimation via Masking:}  
    To recover the triple continuous paramters delay-Doppler-angle in each dimension, the matrix $\mx{U}_i$ is partitioned using masks that exploit the Toeplitz structure along each dimension separately. For instance, in dimension 1 (delay parameter part), two masks $i^{\rm up}_{\tau}$ and $i^{\rm low}_{\tau}$ are defined such that:
    \begin{equation}
        \mathbf{u}^{\mathsf{T}}_{i,1,\text{up}} \triangleq \mathbf{U}(i^{\rm up}_{\tau},:), \quad \mathbf{u}^{\mathsf{T}}_{i,1,\text{low}} \triangleq \mathbf{U}(i^{\rm low}_{\tau},:).
    \end{equation}
    Then, the generalized eigenvalue problem
    \begin{equation}
        \mathbf{u}_{i,1,\text{up}} \mathbf{u}^{\mathsf{T}}_{i,1,\text{low}} \mathbf{z}_{\ell} = \lambda'_{\ell}\, \mathbf{u}_{i,1,\text{up}} \mathbf{u}^{\mathsf{T}}_{i,1,\text{up}} \mathbf{z}_{\ell}
    \end{equation}
    is solved. The delay parameters collected in a vector $\widehat{\bs{\tau}}_i\triangleq [\widehat{\bs{\tau}}_{i,1},...,\widehat{\bs{\tau}}_{i,\hat{s}_i} ]$ are then estimated by
    \begin{equation}
        \widehat{\tau}_{i,\ell} = \operatorname{mod}\left(\tfrac{\operatorname{Im}(\log(\lambda'_{\ell}))}{2\pi}, 1\right), \ell=1,...,\hat{s}_i,
    \end{equation}
    and similar procedures are applied to estimate Doppler $\widehat{\bs{\nu}}_i$ and angles $\widehat{\bs{\theta}}_{i}$.

     \textbf{Triple Pairing:}  
    After obtaining separate estimates $\widehat{{\tau}}_{i,\ell}$, $\widehat{{\nu}}_{i,\ell}$, and $\widehat{{\theta}}_{i,\ell}$, a pairing step is necessary to associate the correct continuous components across the three dimensions. This is accomplished by an exhaustive search over all possible permutations of $\widehat{\bs{\nu}}_{i}$ and $\widehat{\bs{\theta}}_{i}$ such that the combined steering vectors $        \mathbf{a}(\widehat{{\tau}}_{i,\ell},\widehat{{\nu}}_{i,\ell},\widehat{{\theta}}_{i,\ell})
$ maximize a correlation metric with the subspace spanned by $\mathbf{U}_i$. The optimal permutation yields the correctly paired triplets.

     \textbf{Power Estimation:}  
    With the triple parameters properly paired, the amplitudes $p_k$ are estimated by solving a linear least-squares problem. In this step, the model in (\ref{eq:decomp}) is vectorized and the unknown powers are obtained by minimizing the discrepancy between the observed matrix and its Vandermonde representation. This step is also necessary as in the noisy case, estimating powers of the triple components can lead a robust estimate of the total number of involved components i.e. $\hat{s}_i$.

This approach is particularly useful for multidimensional parameter estimation problems, where exploiting the Toeplitz structure leads to efficient and robust parameter recovery.

\subsection{Dual Approach via Dual Polynomials}\label{sec:dual_approach}
Another way to estimate the triple continuous parameters is to form the dual problem of \eqref{eq.primalprob_noisy} given by:

\begin{align}\label{eq.dualprob_noisy}
 &\hspace{-3pt}\max_{\substack{{\mx{q}}\in\mathbb{C}^{L\times 1}\\\zeta_i\in[0,1)^3}}~{\rm Re}~\langle \mx{q}, \mx{y}\rangle-\eta \|\mx{q}\|_2~\\
 &{\rm s.t.}~\Big\|\sum_{n=0}^{L-1}{q}_n\mx{B}_n^i~\mx{a}_{\rm 3D}^*(\zeta^i)\Big\|_2\le 1,~i= 1..., R.
 \end{align}

 According to trigonometric polynomials theory \cite{candes2014towards,dumitrescu2017positive}, the latter optimization problem can be stated equivalently as an SDP problem given by:

\begin{equation}\label{prob.sdp}
    \begin{aligned}
    &\underset{\substack{\mx{q}\in\mathbb{C}^{L\times 1}, \bm{Q}\in \mathbb{C}^{L\times L}}}{\rm max}~~{\rm Re}\big\{\langle \bm{\lambda}, \bm{y}\rangle\big\}\\
    &~~~~{\rm s.t.}~~
    \begin{bmatrix}
        \bm{Q}&\sum_{n=0}^{L-1}{q}^{\ast}_n{\mx{B}_n^i}^{\mathsf{H}}\\
        \sum_{n=0}^{L-1}{q}_n\mx{B}_n^i&\bm{I}_{k_i}
    \end{bmatrix}\succeq \bm{0}, \quad i \in [r],\\
    & ~~~~~~\langle \mathcal{T}({\bm{e}_l}), \bm{Q}  \rangle=1_{l=0}, \quad l=-L+1,..., L-1,
    \end{aligned}
\end{equation}
where $\mathcal{T}(\mx{z})$ for an arbitrary $\mx{z}$ shows the Toeplitz operator whose first row is $\mx{z}$.
After finding the dual vector $\mathbf{q}$ by solving the above SDP problem, we construct a function that is referred to as the dual polynomial function given by:
\begin{align}\label{eq:dual_pol_fun}
 f_i(\zeta)\triangleq  \Bigl\|\sum_{n=0}^{L-1}{q}_n\,\mathbf{B}_n^i\,\mathbf{a}^*(\zeta^i)\Bigr\|_2, \quad i=1,\ldots,R.
\end{align}
The peaks of this dual polynomial $f_i(\zeta)$ within a desired fine 3D grid yield estimates for the triple delay-Doppler-angle parameters \cite{daei2024timely,candes2014towards,fernandez2016super}.

In the next section, we provide collaborative methods to exploit these $R$ functions to improve the estimation of common targets or common communication scatterers.

\section{Collaborative Multi-User Fusion Methods}\label{sec:fusion_methods}
In a multi-user scenario, each user computes its own dual polynomial function $f_i(\zeta)$ and extracts an individual estimate $
\hat{\zeta}_i = \arg\max_{\zeta} f_i(\zeta)$.
However, Since all users observe the same underlying target, key parameters--including the angle-of-arrival (AoA), the distance from the target to the base station, and the target’s relative velocity--are measured repeatedly. Aggregating these $R$ estimates into a single, robust estimate helps to mitigate the effects of noise and outliers. In this work, we consider several fusion techniques:

\textbf{Pointwise Average Fusion}:
The aggregated dual polynomial is computed as the mean of the individual dual polynomials:$  f_{\text{avg}}(\zeta) = \tfrac{1}{R}\sum_{i=1}^{R} f_i(\zeta).$
The common target parameter is then estimated as $\hat{\zeta}_{\text{avg}} = \arg \max_{\zeta} f_{\text{avg}}(\zeta).$

\textbf{Pointwise Maximum Fusion}:
Alternatively, one may compute the aggregated function by taking the pointwise maximum over the users: $   f_{\text{max}}(\zeta) = \max_{i=1,\ldots,R} f_i(\zeta).
$
The corresponding estimate is given by $\hat{\zeta}_{\text{max}} = \arg \max_{\zeta} f_{\text{max}}(\zeta)$.

\textbf{Weighted Average Fusion}:
If some users are more reliable than others, we can assign each user a weight $w_i$, for example based on the peak value of its dual polynomial: $w_i = \max_{\zeta} f_i(\zeta).
$
Then, the weighted aggregated dual polynomial is given by: $   f_{\text{wavg}}(\zeta) = \tfrac{\sum_{i=1}^{R} w_i\, f_i(\zeta)}{\sum_{i=1}^{R} w_i},
$ with the corresponding estimate $\hat{\zeta}_{\text{wavg}} = \arg \max_{\zeta} f_{\text{wavg}}(\zeta).$

\textbf{Aligned Dual Polynomial Fusion}:
In our codebook design, each user's codebook $\mx{D}_i$ is given by $\mathbf{D}_i = c_{\rm user}(i) \, \mathbf{A}_i$
where the complex scalar $c_{\rm user}(i)={\rm e}^{\tfrac{j 2\pi (i-1)}{R}}$ is a distinct code for user $i$ and $\mathbf{A}_i$ is a user-specific dictionary (obtained by applying a circular shift and a small perturbation to a common basis subspace). This user-specific code introduces a phase offset in the dual polynomial for each user. To remove these offsets, we de-phase each user’s dual polynomial by multiplying it by $\tfrac{1}{c_{\rm user}(i)}$, i.e., we define the aligned dual polynomial for user $i$ as $\tilde{f}_i(\zeta)=\tfrac{1}{c_{\rm user}(i)} f_i(\zeta)$.  Then, the aggregated dual polynomial is computed as the mean of the aligned dual polynomials: $  f_{\text{agg}}(\zeta) = \left|\tfrac{1}{R}\sum_{i=1}^{R} \tilde{f}_i(\zeta)\right|^2,
$ and the common target parameter is estimated by $\hat{\zeta}_{\text{agg}} = \arg\max_{\zeta} f_{\text{agg}}(\zeta)
$. This approach removes the user-specific phase differences introduced by the codes and allows the common target parameter to be more clearly revealed.

\subsection{Non-Collaborative Methods}
For comparison, in a non-collaborative approach each user independently estimates the target parameter as $\hat{\zeta}_i = \arg\max_{\zeta} f_i(\zeta),
$
and the final estimate can be obtained by averaging over all users: $\hat{\zeta}_{\text{non-collab}} = \tfrac{1}{R}\sum_{i=1}^{R} \hat{\zeta}_i$. Since individual estimates can be corrupted by noise or outliers, this non-collaborative method usually results in higher parameter estimation error and, consequently, a higher \gf{symbol error rate (SER)} during message recovery.

\subsection{Message Recovery Performance}
After the common target (or scatterer) parameters have been estimated--using either collaborative or non-collaborative methods--the recovered parameter vector $\widehat{\zeta}_{\ell,i} = (\widehat{\tau}_{\ell,i},\widehat{\nu}_{\ell,i},\widehat{\theta}_{\ell,i}), \ell=1,..., \hat{s}_i, i=1,..., R$ is utilized to construct a recovery dictionary for the transmitted message. Specifically, the estimated parameters are employed to form the dictionary, and the transmitted message is then recovered using a least-squares solution. Specifically, by replacing the estimates $\widehat{\zeta}_{k,i}, k=1, \ldots, \widehat{s}_i, i\in[{R}]$ into \eqref{eq:measurement2}, we form the following system of equations:
\begin{align}\label{eq:overdetermined_equations}
\scalebox{.7}{$
         \begin{bmatrix}
              \mx{e}_{0}^{\mathsf{T}}
              \mx{a}_{\rm 3D}(\widehat{\zeta}_{1,1})
              {\mx{d}_{0}^1}^{\mathsf{T}}
              &\hdots&\mx{e}_{0}^{\mathsf{T}}\mx{a}_{\rm 3D}(\widehat{\zeta}_{R,s_R}){\mx{d}_{0}^R}^{\mathsf{T}}\\
              \vdots&\ddots&\vdots\\
           \mx{e}_{L-1}^{\mathsf{T}}
              \mx{a}_{\rm 3D}(\hat{\zeta}_{1,1})
              {\mx{d}_{L-1}^1}^{\mathsf{T}}
             &\hdots&\mx{e}_{L-1}^{\mathsf{T}}\mx{a}_{\rm 3D}(\widehat{\zeta}_{s_R,R}){\mx{d}_{L-1}^R}^{\mathsf{T}}\\
         \end{bmatrix} \begin{bmatrix}
             c_{1,1} \mx{f}_1\\
             \vdots\\
             c_{s_1,1} \mx{f}_1\\
             \vdots\\
             c_{1,R} \mx{f}_r\\
             \vdots\\
             c_{s_R, R} \mx{f}_R\\
         \end{bmatrix}=\begin{bmatrix}
             y_0\\
             \vdots\\
             y_{L-1}
         \end{bmatrix}$}.
     \end{align}
The least square minimizer is the solution of the latter system of equations. By further constraining the signal to have unit norm $\|\mx{f}_i\|_2=1$, we can estimate both message vectors and the channel amplitudes as detailed in \cite[Eq. (29)]{daei2024timely}. This approach leverages accurate parameter estimation to improve overall message recovery performance.

\section{Numerical Results}\label{sec.simulations}
\begin{figure}
    \centering
    \includegraphics[scale=0.2]{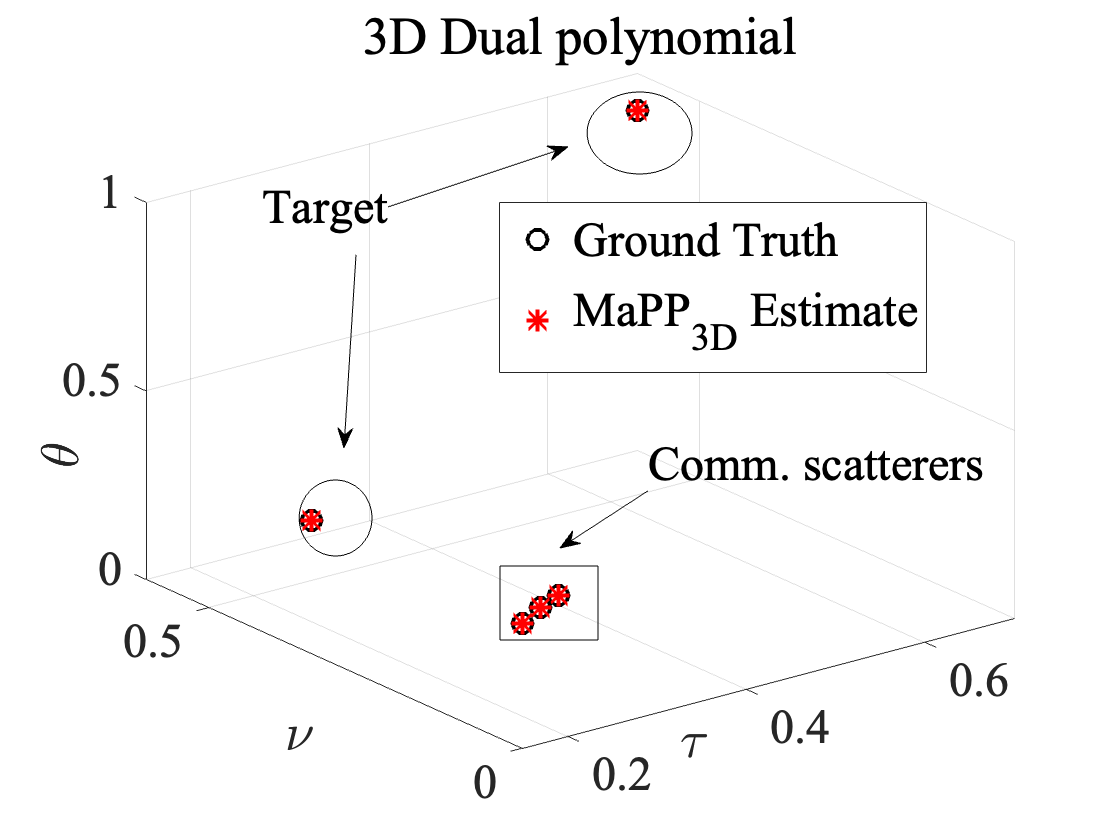}
    \caption{The triple delay-Doppler-angle recovery by ${\rm Mapp}_{\rm 3D}$ method provided in Section \ref{sec:mapp_3d}. The communication scatterers form a cluster and exhibit negligible Doppler shifts while the targets have varying levels of Doppler shifts.}
    \label{fig:3D_recovery}
\end{figure}
\begin{figure}
    \centering
    \includegraphics[scale=0.2]{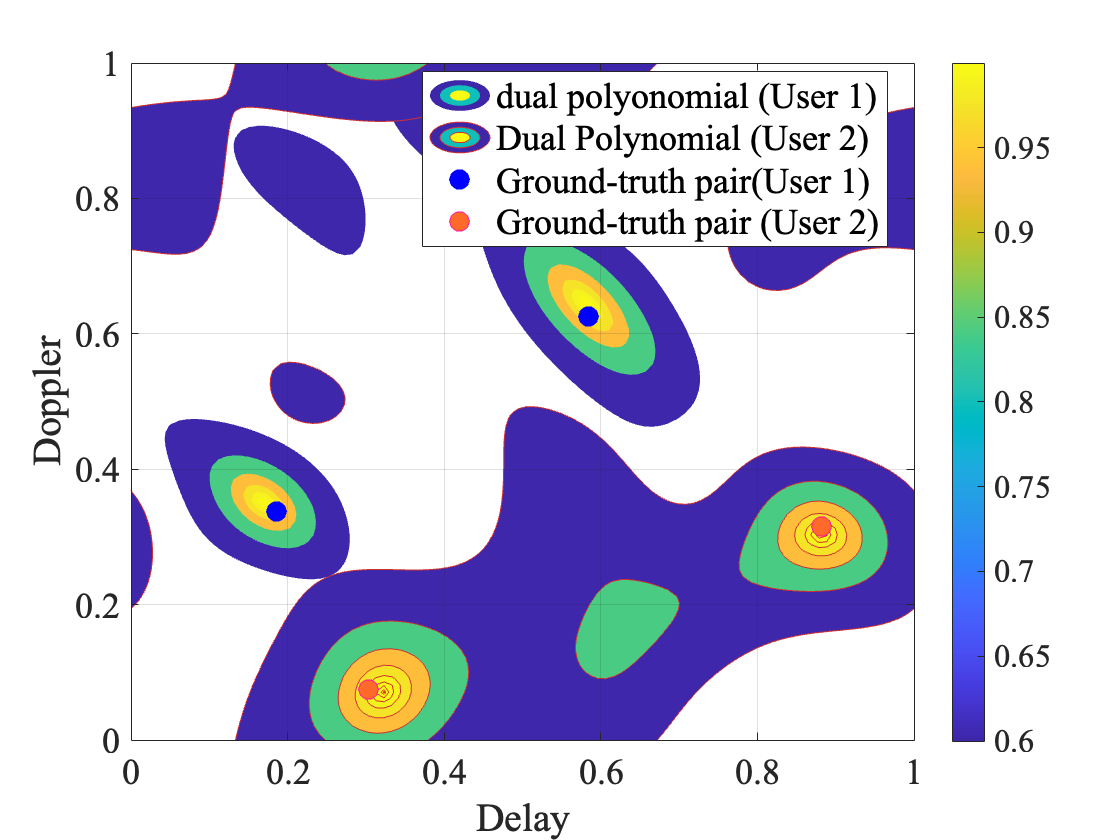}
    \caption{Delay-Doppler estimation by dual polynomial functions. There are two users and each has its own dual polynomial function. The x-axis is delay domain while the y-axis is the Doppler domain. The dual polynomial value is shown by color. The SNR is set to $0~dB$.}
    \label{fig:2d_recovery}
\end{figure}
\begin{figure}[h]
    \centering
    
    \begin{subfigure}[]{0.15\textwidth}
        \centering
 \includegraphics[scale=.12,trim={0cm 0cm 0cm 0cm}]{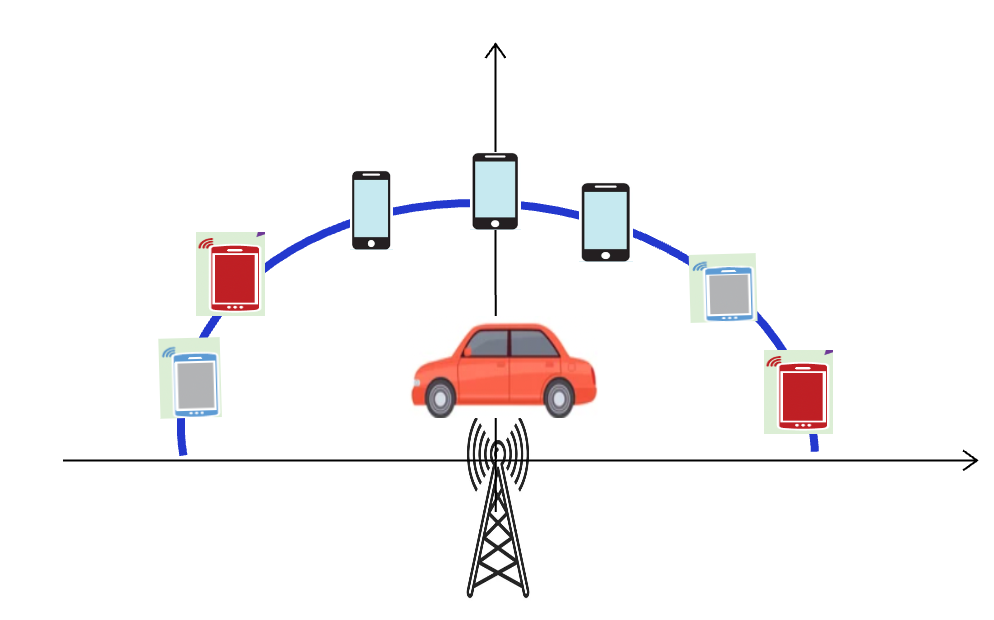}
        \caption{}
        \label{fig:Schematic(a)}
    \end{subfigure}
       \hfill
    \begin{subfigure}[]{0.3\textwidth}
        \centering
 \includegraphics[scale=.26,trim={0cm 0cm 0cm 0cm}]{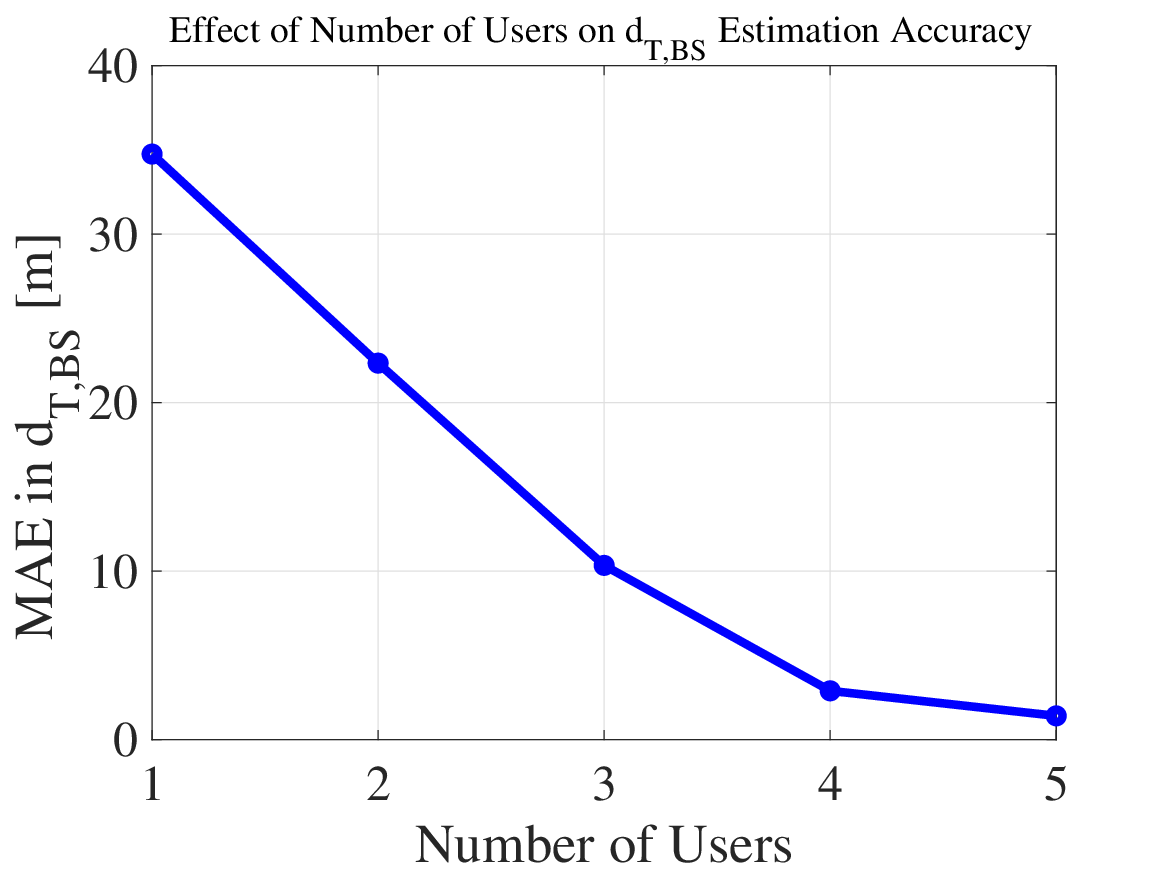}
        \caption{}
      \label{fig:colab_performance(b)}
    \end{subfigure}

    \caption{\ref{fig:Schematic(a)}: Multi-user collaboration to sense the common target and scatterers. Multiple users send their signals towards the ISAC BS. All signals hit the common target/scatterers and the echoes are received at the BS side. \ref{fig:colab_performance(b)}: The estimation of target distance to BS using delay measurements of multiple users}
     \label{fig:capacity_bounds}
\end{figure}
\begin{figure}[h]
    \centering
    
    \begin{subfigure}[]{0.24\textwidth}
        \centering
 \includegraphics[scale=.23,trim={0cm 0cm 0cm 0cm}]{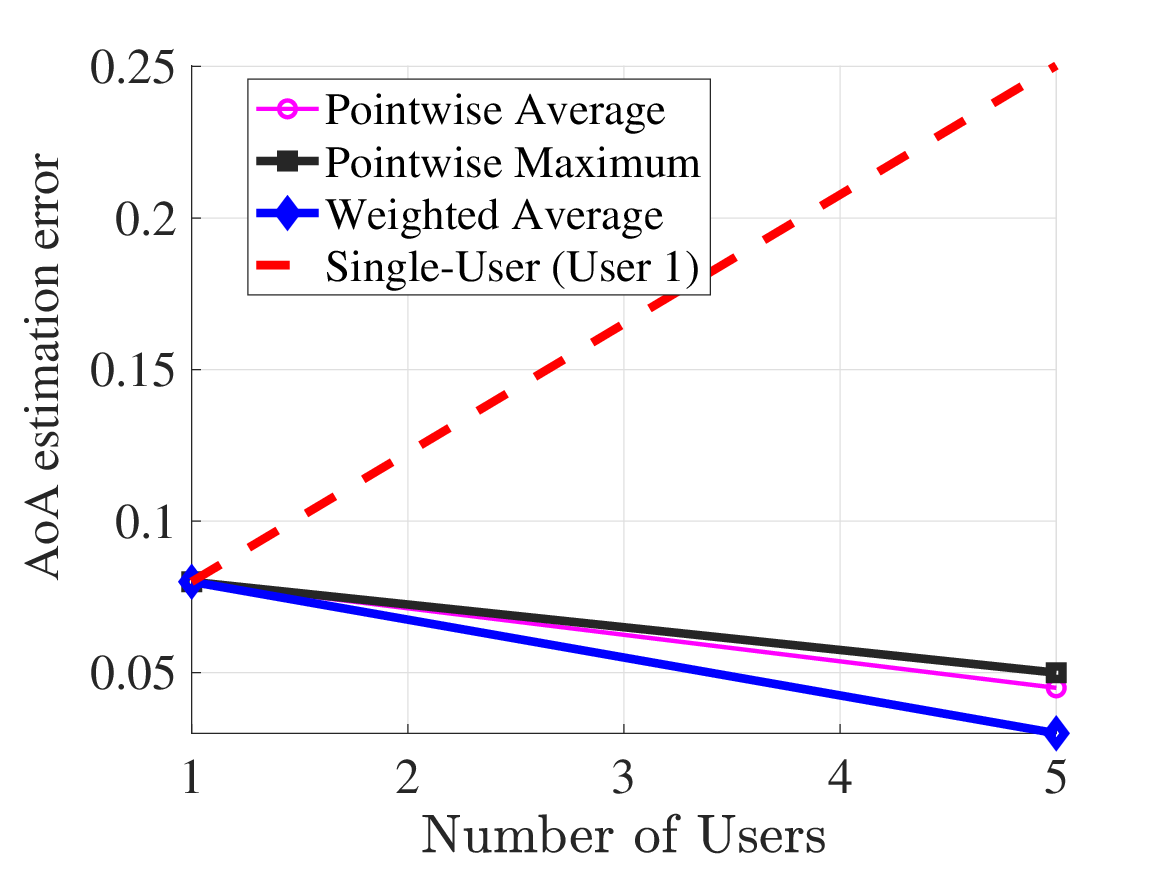}
        \caption{}
        \label{fig:colab1}
    \end{subfigure}
       \hfill
    \begin{subfigure}[]{0.24\textwidth}
        \centering
 \includegraphics[scale=.23,trim={0cm 0cm 0cm 0cm}]{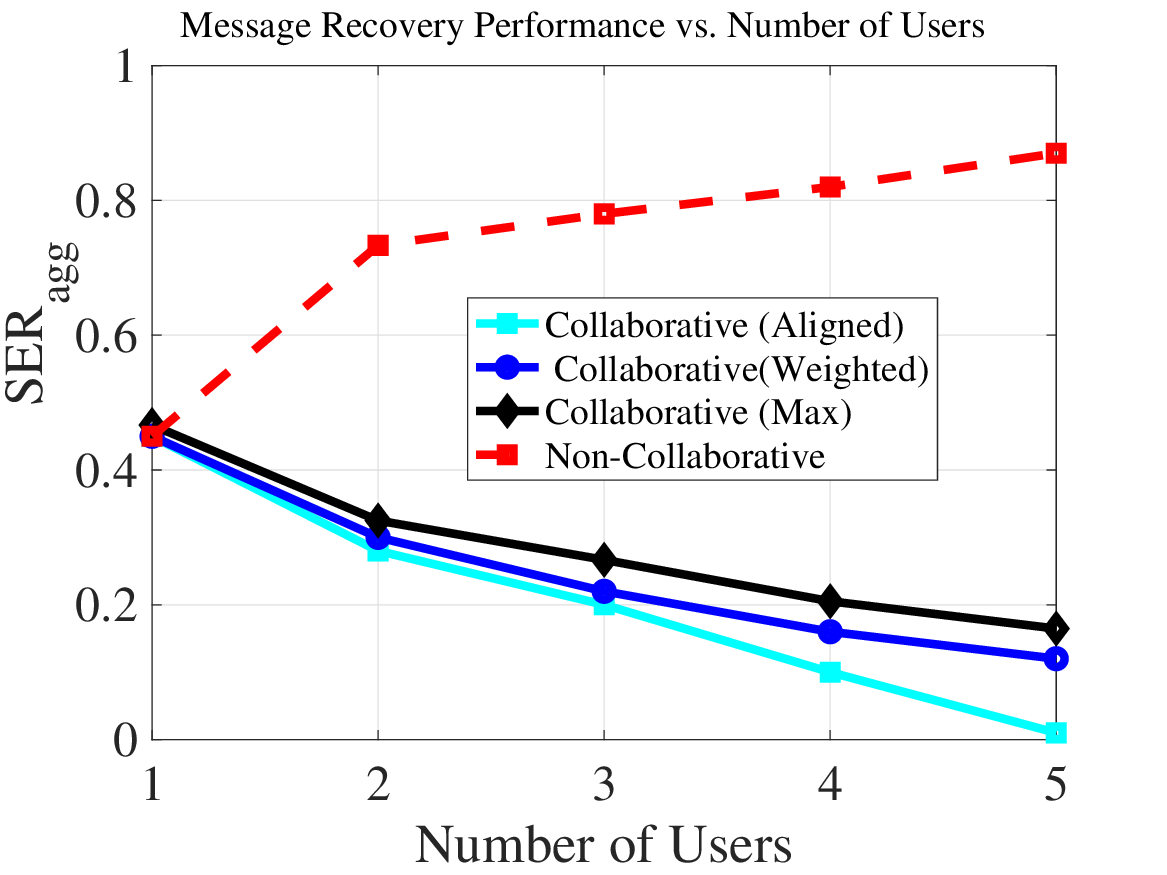}
        \caption{}
      \label{fig:colab2}
    \end{subfigure}

    \caption{Comparison of collaborative methods with non-collaborative methods.  (\ref{fig:colab1}): This figure shows the AoA estimation error for individual AoA estimates and collaborative estimates the AoA corresponding to the common target. (\ref{fig:colab2}): This figure compares the message recovery performance between non-collaborative methods and collaborative methods including weighted average, average, pointwise maximum aligned dual polynomial. The received signal is highly noisy with ${\rm SNR}=0dB$.}
     \label{fig:colab}
\end{figure}
In this section, we provide several numerical simulation results that evaluate the performance of our proposed approach. The simulation results are divided into multiple subsections.

\subsection{Continuous Parameter Recovery}
In Figure \ref{fig:3D_recovery}, we have designed an experiment for recovery of the triple delay-Doppler-angle from the noisy measurements provided in \eqref{eq:measurement2}. The signal to noise ratio (SNR) in decibels is defined as follows:

\begin{equation}
\text{SNR}_{\text{dB}} = 10\log_{10}\left(\tfrac{\|\mx{y}\|_2^2}{L\sigma^2}\right),
\end{equation}
where \(\|\mx{y}-\bs{\epsilon}\|_2^2\) denotes the total power of the noiseless observation. In this experiment, we set the SNR to be $30 dB$ and set $P=Q=N_r=6$ to take $L=PQN_r=6^3$ time-freq-space samples at the ISAC receiver.
Then, we choose the dual approach of Section \ref{sec:dual_approach} by setting $\eta\triangleq\sigma \sqrt{L+\sqrt{2L\log(L)}} $ which is inspired by \cite{daei2024timely,daei2023blind} and obtain the dual polynomial function for User 1 in Figure \ref{fig:3D_recovery}.
As is mentioned in this figure, communication scatterers form a cluster of parameters with negligible Doppler shifts while targets are assumed to have varying levels of Doppler shifts. In this way, the communications scatterers and radar targets can be distinguished in 3D delay-Doppler-angle space. In another experiment shown in Figure \ref{fig:2d_recovery}, we consider two users with different polynomials and $s_1=s_2=2$. The dual polynomial values are depicted by colors (yellow represent one and blue shows zero). The SNR is set to be $0dB$ which is a very noisy regime. Despite this high-noise settings, the proposed approach succeeds to find the ground-truth delay-Doppler pairs.   

\subsection{Collaboration in Target Localization}
We consider a BS located at $\mx{x}_{BS}=[0, 0]^{\mathsf{T}}$ and a target located at $\mx{x}_{T}=[50, 30]^{\mathsf{T}}$ as is depicted in Figure \ref{fig:Schematic(a)}. The goal is to estimate $d_{T,BS}\triangleq \|\mx{x}_T - \mx{x}_{BS}\|_2$ via time measurements i.e. $Q$ and $N_r$ are assumed to be one for simplicity in a noisy setup with ${\rm SNR}_{\rm dB}=5 dB$. Each user senses a common target, and the corresponding propagation delay is modeled as
\begin{equation}
\overline{\tau}_{i,1} = \tfrac{\|\mx{x}_i - \mx{x}_T\|_2 + \|\mx{x}_T - \mx{x}_{BS}\|_2}{c_{\rm light}},
\end{equation}
where \(\mx{x}_i\) denotes the location of the \(i\)th user and \(c_{\rm light}\triangleq 3\times10^8\) is the speed of light.
Since the physical delays in typical scenarios are very small compared to the system’s time scale i.e. $T$, an linear mapping is applied to normalize the delays into the interval \([0,1]\):
\begin{equation}
\tau_{i,1} = \tfrac{\overline{\tau}_{i,1} - \min_i \overline{\tau}_{i,1}}{\max_i\overline{\tau}_{i,1} - \min_i \overline{\tau}_{i,1}}.
\end{equation}
This mapping ensures that the super-resolution method provided in Section \ref{sec:mapp_3d} has sufficient dynamic range for delay separation. We solve the SDP problem \eqref{eq:3DfinalSDP} and then we find $R$ delay estimates of the common target. The estimated normalized delay values are then mapped back to the physical delay domain to get $\widehat{\overline{\tau}}_i, i=1,..., R$.
The measured distance-sum for the \(i\)th user is given by
\begin{equation}
d_{\mathrm{meas}}(i) = c_{\rm light}\,\widehat{\overline{\tau}}_i,
\end{equation}
where \(\widehat{\overline{\tau}}_i\) is the estimate of the delay parameter by the $i$-th user. These measurements are used in the following nonlinear least-squares problem to estimate the target location \(x_T\):
\begin{equation}
\widehat{\mx{x}}_T
=
\mathop{\arg\min}_{\mx{x}_T}  \sum_{i=1}^{R} \Big( \|\mx{x}_i - \mx{x}_T\|_2 + \|\mx{x}_T - \mx{x}_{BS}\|_2 - d_{\mathrm{meas}}(i) \Big)^2.
\end{equation}
Finally, the estimated target-to-BS distance is obtained as $\widehat{d}_{T,BS} \triangleq \|\widehat{\mx{x}}_T - \mx{x}_{BS}\|$.
To assess the benefit of collaboration, we conduct Monte Carlo simulations over a range of user numbers \(R\) (from $1$ to $5$). For each \(R\), the mean absolute error (MAE) in \(d_{T,BS}\) is computed:
\begin{equation}
\text{MAE} = \tfrac{1}{N_{\text{trial}}} \sum_{j=1}^{N_{\text{trial}}} \left| \widehat{d}_{T,BS}^{(j)} - d_{T,BS} \right|,
\end{equation}
where \(d_{T,BS}\triangleq \|\mx{x}_T - \mx{x}_{BS}\|_2\).

Figure~\ref{fig:colab_performance(b)} shows that the estimation error is high when only one user is used, due to the inherent ambiguity in the localization problem. However, as the number of users increases, the diversity in delay estimates, coming from multiple users, improves the constraints in the localization problem, thereby significantly reducing the error in estimating \(d_{T,BS}\). 
\subsection{Collaborative AoA Estimation}
In this section, we have done an experiment to examine the performance of our proposed collaborative approaches in Section \ref{sec:fusion_methods} in estimating the AoA of the common target, viewed by different users located at different places. Unlike delay parameters that must be different for different users since they are located at different places, the AoA corresponding to different user must be similar as it shows the AoA from the target to the BS. In this case, we consider a noisy observations with $\text{SNR}=0\text{~dB}$. In Figure \ref{fig:colab}, we investigate the effect of collaboration on the AoA estimation performance. In Figure \ref{fig:colab1}, the AoA estimation error is depicted. As it turns out, relying on the user-specific AoA estimates leads to larger errors compared to collaborative estimates obtained by pointwise averaged, weighted averaged and pointwise maximum methods provided in Section \ref{sec:fusion_methods}. Moreover, in Figure \ref{fig:colab2}, the message recovery performance is evaluated based on the aggregate {symbol error rate (SER)} defined as
\begin{align}
 {\rm SER}_{\rm agg}\triangleq \tfrac{\sum_{i=1}^R k_i {\rm SER}_i}{\sum_{i=1}^R k_i}   
\end{align}
where ${\rm SER}_i$ is the SER for user $i$. In this experiment, we set $k_i=3,\forall i=1,..., R$ and ${\rm SNR}=0dB$. Transmissions employ an 8-ary amplitude-shift-keying (8-ASK) constellation and we consider one common target with only AoA measurements ($P=Q=1$ and $L=N_r=30$).  As shown in Figure \ref{fig:colab2}, collaborative methods can even lead to enhanced message recovery when they collaborate to sense the common targets and scatterers. 

\subsection{Discussion}
The simulation results highlight several key points:
\begin{itemize}
    \item \textbf{Geometric Diversity:} A single delay estimation obtained by a single dual polynomial yields an ambiguous estimate of the target range, velocity and angle. In contrast, when multiple users are involved, their diverse spatial positions and velocities provide multiple independent estimates, effectively reducing localization ambiguity.
    \item \textbf{Delay Normalization:} The affine mapping of physical delays to the interval \([0,1]\) is crucial for ensuring a sufficient spread for super-resolution. This processing step is essential when the natural spread of delays is very small relative to the chosen time scale.
    \item \textbf{Estimation Accuracy:} The average error in estimating \(d_{T,BS}\) decreases as the number of users increases. This confirms the advantage of collaborative sensing, where the fusion of multiple dual polynomials leads to improved accuracy.
\end{itemize}


\section{Conclusion and Future Directions}\label{sec.conclusion}

In this paper, we introduced a novel collaborative sensing and communication strategy for ISAC uplink systems, where multiple ISAC-enabled users transmit data to a single ISAC-enabled receiver. Unlike traditional multi-user communication frameworks, which treat interference as a limiting factor, our approach exploits the inherent diversity in both spatial positioning and velocity differences among users to enhance both target sensing and data decoding.
To achieve this, we formulated a 3D super-resolution framework that accurately distinguishes communication scatterers from radar targets by jointly estimating their delay, Doppler shift, and AoA parameters. By leveraging the observation that communication scatterers tend to form stationary clusters with negligible Doppler, our method effectively isolates dynamic targets, enabling precise localization. Furthermore, we developed multiple collaborative estimation schemes that fuse multi-user measurements, significantly boosting sensing resolution and data recovery performance.
Extensive numerical evaluations demonstrated that exploiting both spatial and velocity diversity leads to substantial performance gains over conventional multi-user uplink systems that neglect these rich sources of diversity. Our results confirm that by jointly optimizing sensing and communication, ISAC uplinks can transform user transmissions into valuable sensing opportunities, rather than treating them as interference. This paradigm shift highlights the potential of ISAC to power the next generation of wireless networks, where seamless environmental awareness and ultra-reliable data transmission are jointly realized.  Future directions include extending our framework to account for dynamic user mobility, real-time implementation challenges, and further optimization of collaborative fusion techniques in extreme noise conditions.


\bibliographystyle{ieeetr}
\bibliography{Ref}
\end{document}